\documentclass[twocolumn,showpacs,preprintnumbers,amsmath,amssymb]{revtex4}

\usepackage{graphicx}

\usepackage{dcolumn}
\usepackage{bm}

\begin{document}
\bibliographystyle{h-physrev3}


\title{Simultaneous Measurement of Ionization and Scintillation from Nuclear Recoils in Liquid Xenon as Target for a Dark Matter Experiment}

\author{E.\,Aprile${}^1$}
\email{age@astro.columbia.edu}
\author{C.\,E.\,Dahl${}^4$}
\author{L.\,DeViveiros${}^3$}
\author{R.\,Gaitskell${}^3$}
\author{K.\,L.\,Giboni${}^1$}
\author{J.\,Kwong${}^4$}
\author{P.\,Majewski${}^1$}
\author{K.\,Ni${}^1$}
\author{T.\,Shutt${}^2$}
\email{tshutt@cwru.edu}
\author{M.\,Yamashita${}^1$}

\affiliation{%
${}^1$ Department of Physics, Columbia University, New York, NY 10027, USA
}%
\affiliation{%
${}^2$Department of Physics, Case Western Reserve University, Cleveland, OH  44106, USA
}%
\affiliation{%
${}^3$Department of Physics, Brown University, Providence, RI 02912, USA
}%
\affiliation{%
${}^4$Department of Physics, Princeton University, Princeton, NJ  08540, USA
}%

\date{\today}

\begin{abstract}

We report the first measurements of the absolute ionization yield of
nuclear recoils in liquid xenon, as a function of energy and electric-field.
Independent experiments were carried out with two dual-phase time projection
chamber prototypes, developed for the XENON Dark Matter project.  We find
that the charge yield increases with decreasing recoil energy, and exhibits
only a weak field dependence. These results are a first
demonstration of the capability of  dual phase xenon detectors  to discriminate
between electron and nuclear recoils, a key requirement for a sensitive dark
matter search at recoil energies down to 20~keV.

\end{abstract}

\maketitle

\textbf{Introduction:}
Current evidence indicates that one quarter of the mass-energy density of the universe is composed of cold, non-baryonic dark matter, which has thus far been observed only through its gravitational interactions with normal matter. Its precise nature is undetermined, but weakly interacting massive particles (WIMPs) are an attractive candidate which may be detectable via rare elastic scattering interactions depositing a few tens of keV in target nuclei.  For a review of the motivation for WIMPs, and current experimental WIMP searches, see \cite{Gaitskell:2004dmrev}. The best limits, from CDMS, are $<$0.06~events/kg/day (in Ge) \cite{Akerib:2005cdmslimit}.  Improving the search sensitivity will require both larger detectors and lower radioactive backgrounds.  An attractive method to achieve this uses liquid noble gas-based detectors, which promise to be readily scalable to the multi-ton scale.  For the XENON experiment \cite{Aprile:2002xenon}, we are developing a large volume liquid xenon (LXe) dual-phase time-projection-chamber (TPC) that features simultaneous measurement of recoil ionization and scintillation  to determine the energy and 3-D localization on an event-by-event basis.  
Nuclear recoils from WIMPs (and neutrons) have denser tracks, and thus have been assumed to have greater electron-ion recombination than electron recoils, providing a basis for discrimination against radioactive background gammas and betas.  However, ionization from these nuclear recoils has not been measured until now.  In this report, we describe the first measurements of the ionization yield of low energy nuclear recoils in LXe  and confirm the baseline discrimination capability of this technique.

Independent measurements were carried out with two separate dual-phase
detectors at Case and Columbia. The active LXe is defined by two
electrodes, a cathode and a gate grid, at a distance of 1.9 cm
(Columbia) and 1.0 cm (Case). An event in the active LXe produces both
ionization electrons  and scintillation photons. If trapping by
impurities is negligible, the electrons which escape initial
recombination drift freely under an applied electric field of a few
kV/cm up to the liquid-gas phase interface where they are extracted
and accelerated by the high ($\sim$10~kV/cm) field applied across the
gas gap.  These accelerated electrons produce proportional
scintillation photons before being collected on the anode. The fields
are created by setting potentials on four electrodes: cathode, gate
grid, anode, and top grid. The electrodes are stretched wire grids,
except for a solid stainless steel cathode in the Case detector and a mesh  top electrode in the Columbia detector.  The wire grids are made with 120~$\mu$m diameter BeCu wires on a 2 mm pitch, except for 40~$\mu$m gate grid in the Case detector, and are soldered to SS (Columbia) and Cirlex (Case) frames. 

Photomultiplier tubes (PMTs) detect both the direct scintillation
light (S1) produced at the event site in the liquid and the
proportional scintillation light (S2) produced in the gas. The time
difference between the two signals,  determines the depth of the
event. The 5 cm diameter, 4 cm long metal channel PMTs (Hamamatsu Model R9288) were developed to be directly coupled with LXe, and have a quantum efficiency of $\sim$15\% at the 178 nm wavelength of the Xe  scintillation \cite{Jortner:1965qe}.  PTFE is used as a VUV reflector enclosing the active LXe volume in both detectors. The Columbia detector uses two PMTs, one in gas and one in liquid.   Because of total internal reflection at the liquid surface (index of refraction 1.65 at 187 K  \cite{Solovov:2003indexscatteringxenon}), the light collection is significantly better for the PMT in the liquid. The total S1 light collection efficiency (fraction of photons incident on the photocathode of  a PMT) is estimated from Monte Carlo to be more than 50\%, uniform across the volume to within 5\%. An internal blue light LED  is used to calibrate and monitor the gain of the two PMTs.   The Case detector uses a single PMT in the gas, with an S1 light collection efficiency of $\sim$16\%.   The PMT gain is monitored using the S1 light from 5.3~MeV alphas from a source described below.   The operation of a dual phase LXe detector is described in more detail in \cite{Aprile:2004proplight}. 

Both detectors use a liquid nitrogen-cooled cold finger cryostat with
temperature control to maintain a stable liquid temperature.  The
detectors were operated with vapor pressures between 2.0 and 2.8 atm,
with better than 1\% stability.  The Case detector also uses a
triple-parallel plate capacitor system to align the liquid surface
with the grids and monitor the level and stability of the liquid
surface, which was\cite{Shutt:2006capmeter} stable to within 20~$\mu$m
over two months.  The LXe in both detectors was purified using a high
temperature getter. The Columbia detector used the purification system
with continuous gas circulation through the getter, developed for the first XENON prototype \cite{Aprile:2004ucla, Aprile:2005progress}.  The Case detector used a similar recirculating  system only at the start of  its two-month run.

External gamma ray sources, including $^{57}$Co, $^{133}$Ba, and $^{137}$Cs, were used for calibration of both detectors.  The Case detector also had an internal ${}^{210}$Po source deposited on the center of the cathode plate, providing 5.3~MeV alpha particles and 100~keV \(^{206}\)Pb nuclear recoils. Events  $<$0.5 mm above the cathode were tagged as possible alphas and \(^{206}\)Pb recoils.  For neutron data, a 5~Ci AmBe source (Columbia) and 25~\(\mu\)Ci \(^{252}\)Cf source (Case) were used, with  lead shielding to attenuate gammas from the sources.

The PMT signals were digitized with multiple ADCs sampling at 5~MHz-1~GHz. The effective trigger threshold for the Case detector was 4.5 electrons for the S2 signal, and 50\% single photoelectron (spe) acceptance for S1. For the Columbia detector the trigger was either the coincidence of the S1 signals from the two PMTs (at a few spe level) or the S1 signal from the PMT in the liquid (at a $\sim$6 spe level).

\begin{figure}[htbp]
\includegraphics[height=5cm]{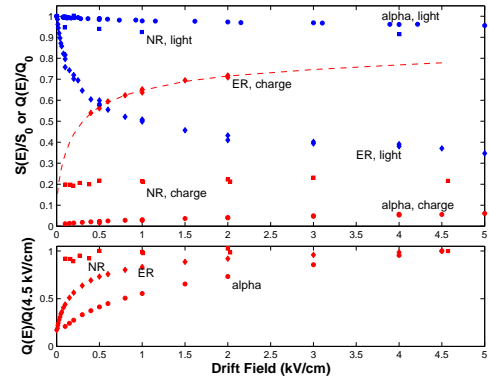}
\caption{\label{yield_vs_field}Top: Field dependence of scintillation and ionization yield in liquid xenon for 122 keV electron recoils (ER), 56 keVr nuclear recoils (NR) and alphas. Bottom:  Ionization yields scaled by their 4.5~kV/cm values.}
\end{figure}

\textbf{Calibration and Raw Data:}
Both detectors were calibrated primarily with ${}^{57}$Co 122~keV
gammas, for both the S1 and S2 signals.  The S1 signal serves as an
energy axis for nuclear recoils, given by
$E_r=E_e/\mathcal{L}_{eff}\cdot S_e/S_r$ where $E_r$ is the nuclear
recoil energy, $E_e$ is a linear electron recoil energy scale based on
the $^{57}$Co S1 peak, $\mathcal{L}_{eff}$ is the effective Lindhard
factor relating the scintillation yields of nuclear and electron
recoils at zero field, and $S_x$ is the loss of scintillation light
due to the recombination suppression by the electric field for species
$x$, equal to $S(\textbf{E})/S_0$ in Fig. \ref{yield_vs_field}.
Recent measurements at Columbia give $S_r$ for 56.5~keV recoils and
$\mathcal{L}_{eff}$ at energies up to 56.5~keV
\cite{Aprile:2005xenonscint}. Other experiments
\cite{Akimov:2001ukdmscint,Arneodo:2000icarusscint,Bernabei:96,Chepel:2005coimbrascint}
have measured $\mathcal{L}_{eff}$ at higher energies.  The Columbia
and other data, except \cite{Bernabei:96}, were parameterized by  $\mathcal{L}_{eff}=0.0984{E_r}^{0.169}$. For simplicity, we assume $S_r$ has no energy dependence since the Columbia measured value is close to unity, as seen in Fig. \ref{yield_vs_field}.   Also shown in Fig. \ref{yield_vs_field} are measurements of $S_e$ as a function of drift field.  Hereafter, nuclear recoil energies calculated in this way will be denoted 'keVr', and electron recoil energies based on the linear S1 scale will be denoted 'keVee'.  In the Case and Columbia detectors the S1 calibration gives, respectively, 1.5~photoelectrons~(pe)/keVee and 5~pe/keVee, at zero electric field, which correspond to 0.28 pe/keVr and 1 pe/keVr for 55 keVr nuclear recoils.

The S2 signal measures the ionization from each event. There is no published data of charge yield vs field for low energy gammas in LXe, so this measurement was performed with ${}^{57}$Co 122~keV gammas, in both the Case \cite{Shutt:2006calibration} and Columbia \cite{Aprile:2006chargelight} detectors operated in single (liquid) phase. The results are shown in Fig. \ref{yield_vs_field}.  The S2 calibration is a function of gas pressure and electric field across the gap \cite{Bolozdynya:1999dualphase}, with the added complication of possible electron multiplication near the anode wires. Typical values  were 19~pe/e$^-$ with 4.6 kV/cm (Case) and 8.4~pe/e$^-$ with 2 kV/cm (Columbia).

\begin{figure}
\includegraphics[height=5.2cm]{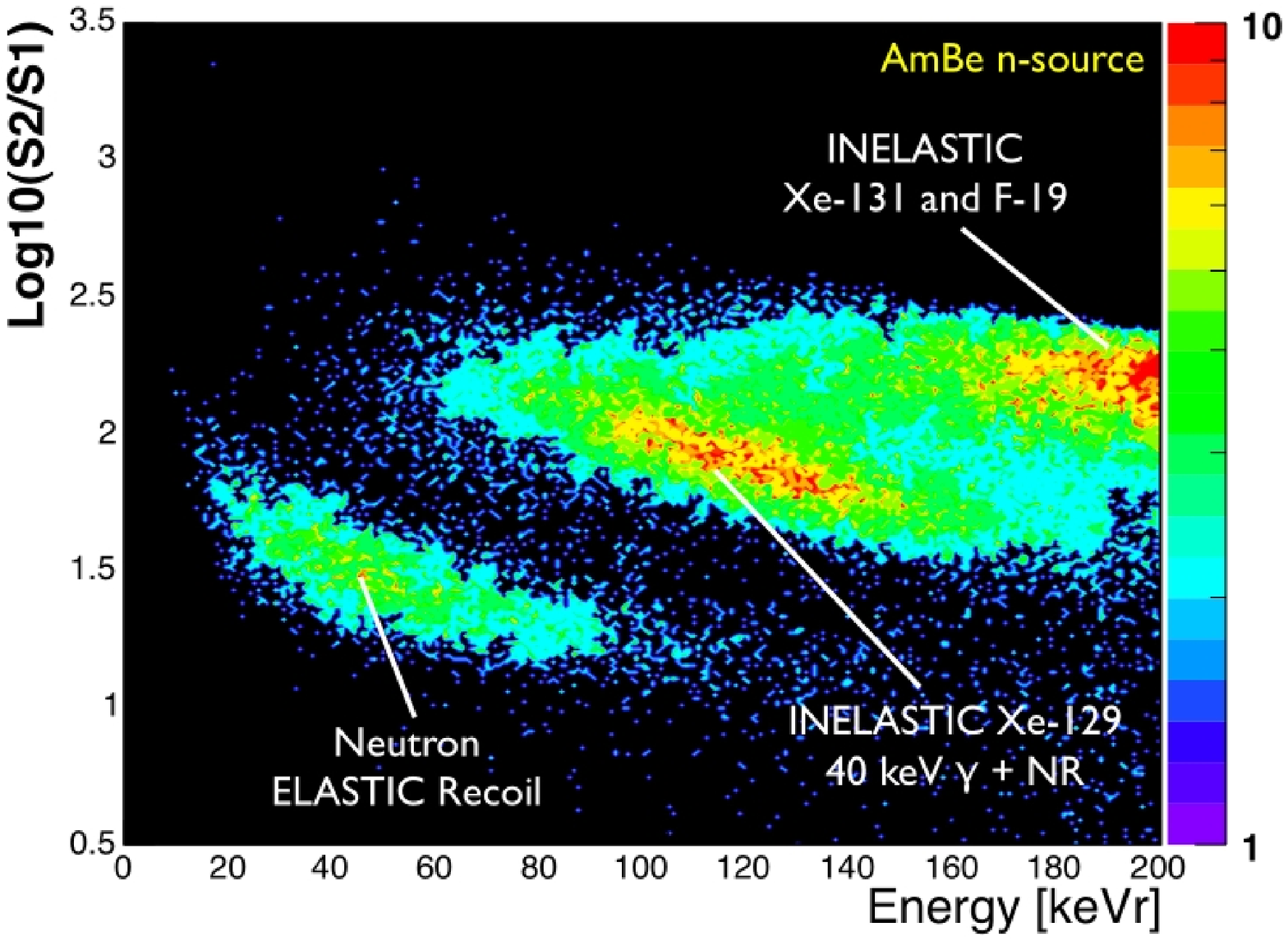}
\includegraphics[height=5.2cm]{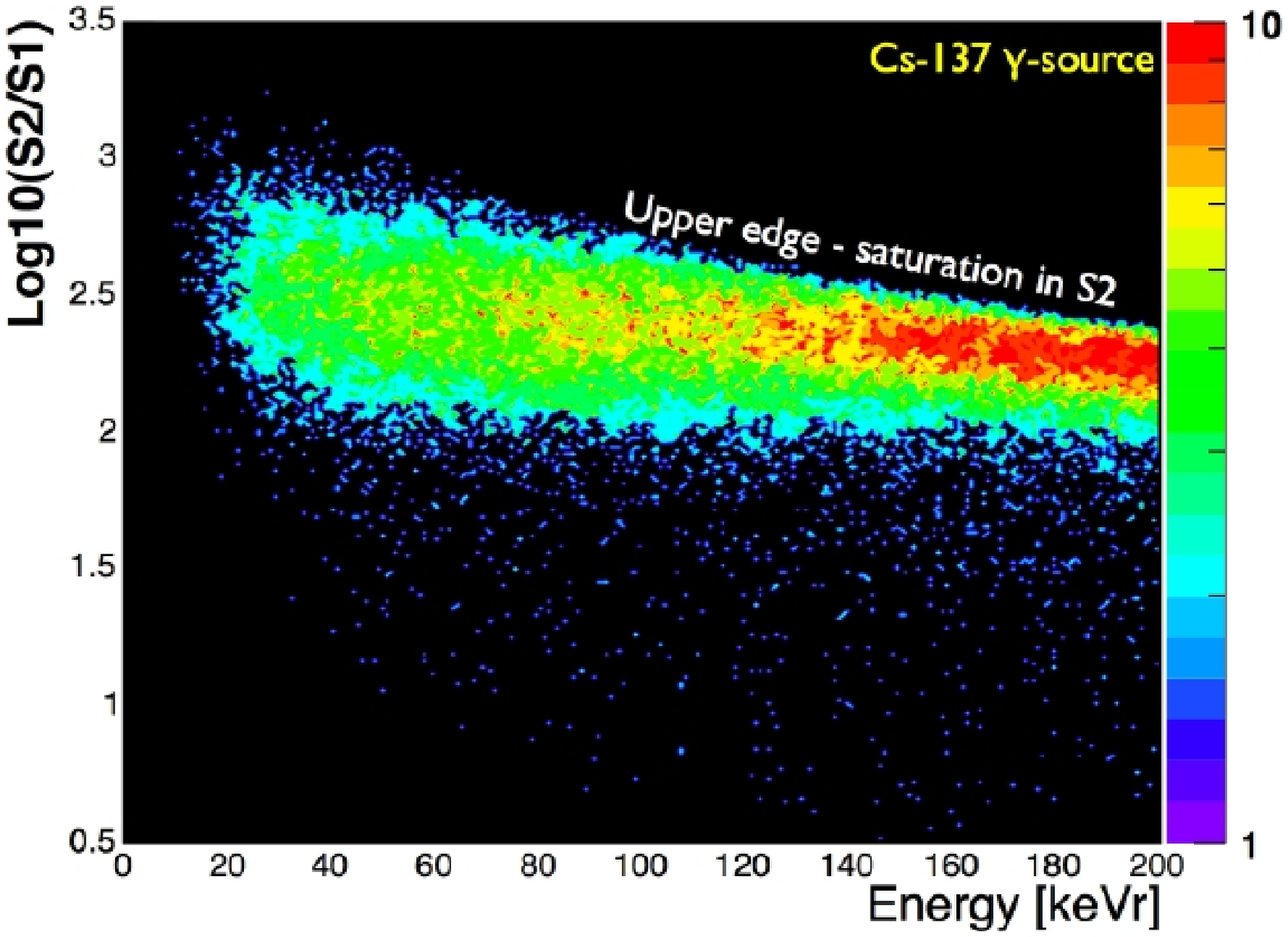}
\caption{\label{2Dplots} Columbia detector response to AmBe neutron (top) and $^{137}$Cs gamma sources (bottom), at 2 kV/cm drift field.}
\end{figure}

\begin{figure}
\includegraphics[height=4.3cm]{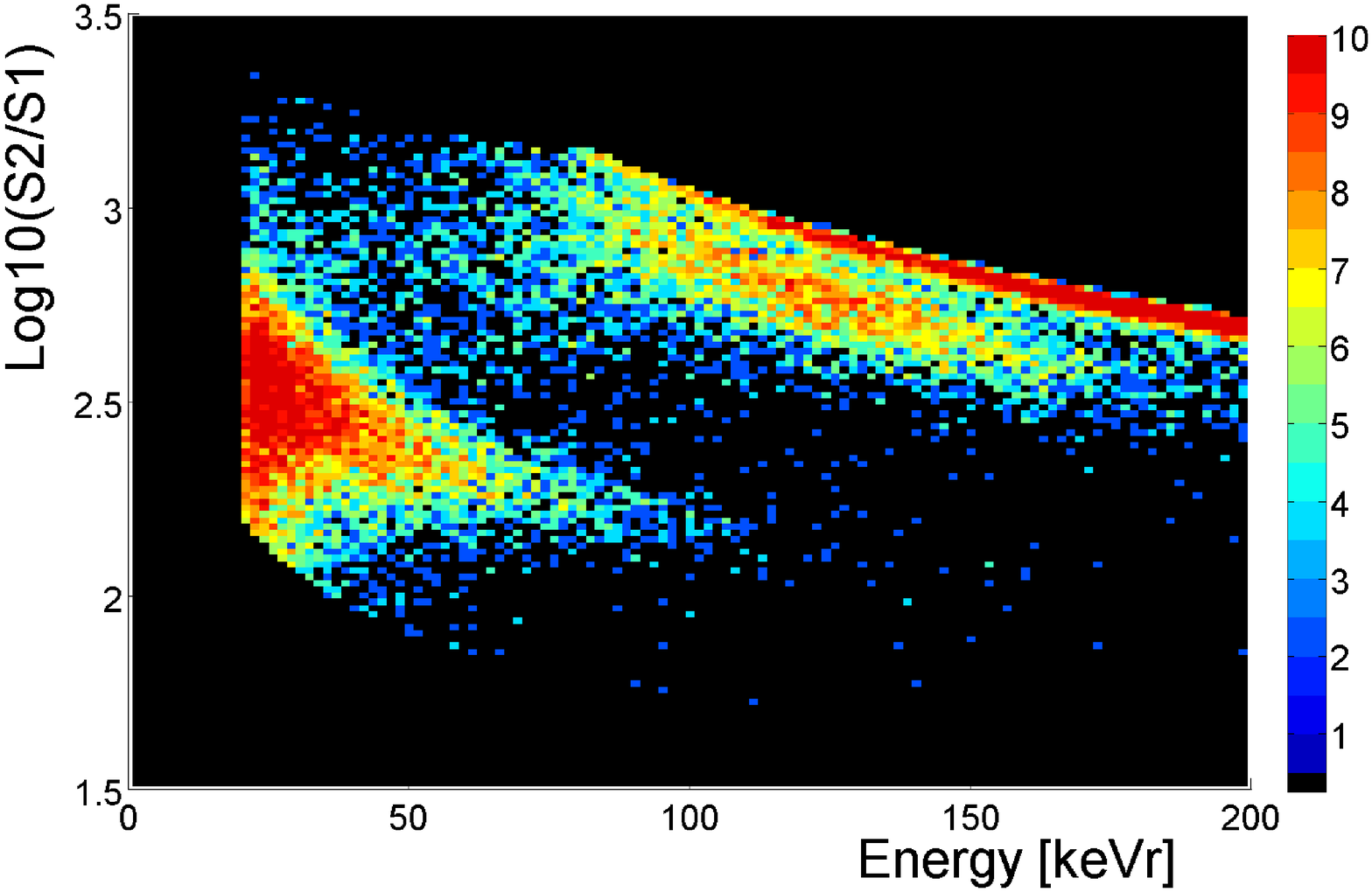}
\includegraphics[height=4.3cm]{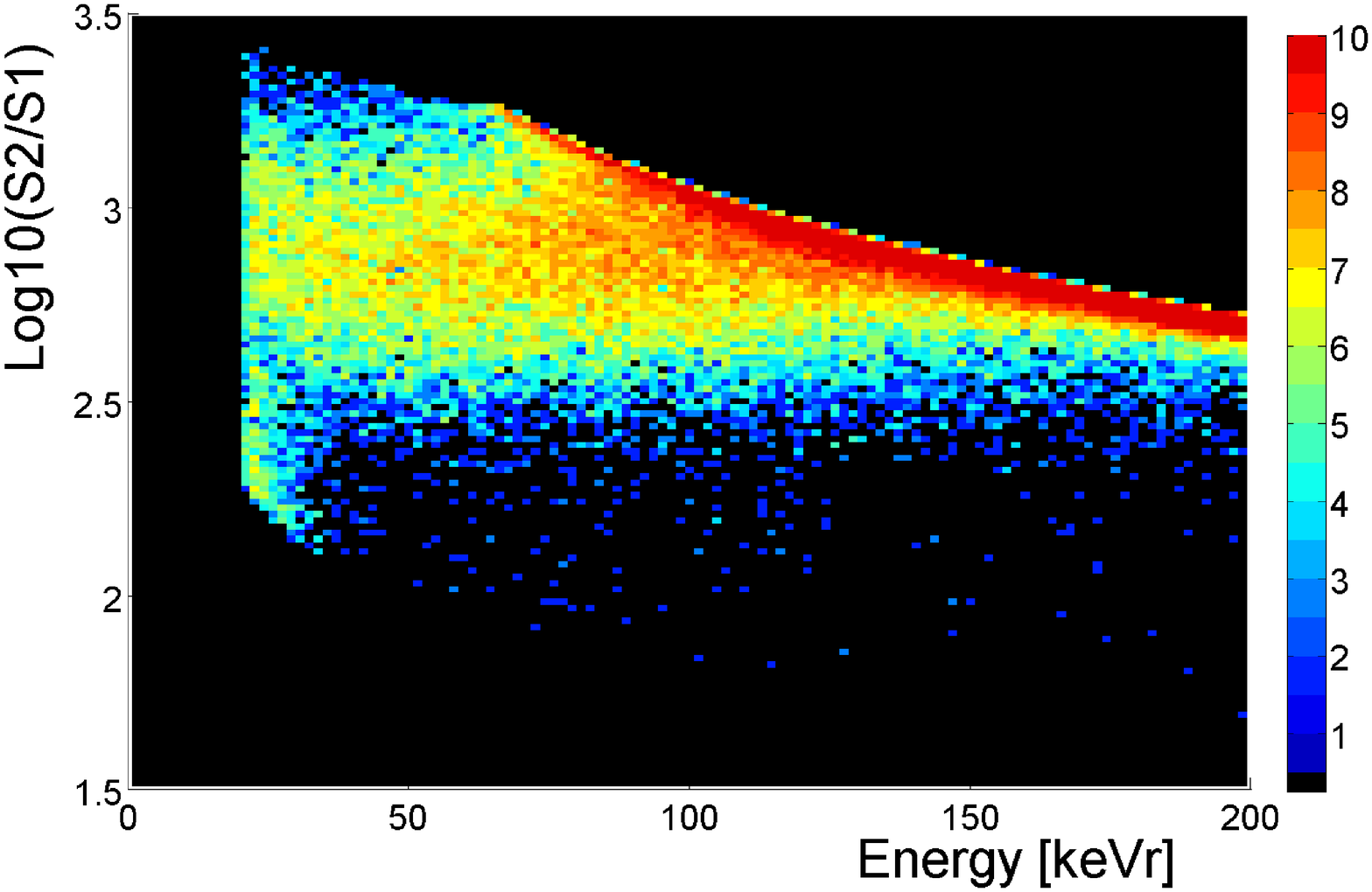}
\caption{\label{2Dcase} Case detector response to \(^{252}\)Cf neutron (top) and $^{133}$Ba gamma sources (bottom) at 1.0 kV/cm drift field. }
\end{figure}

Figures \ref{2Dplots} and \ref{2Dcase} shows the detectors' responses to neutron and low energy Compton scattering events. The logarithm of the ratio S2/S1 is plotted as a function of nuclear recoil energy (keVr). In both detectors, the elastic nuclear recoil band is clearly separated from the electron recoil events. The events identified around 40 keVee are from gamma rays produced by neutron inelastic scattering on $^{129}$Xe. Additional gamma ray events are emitted following inelastic scattering of neutrons on $^{131}$Xe and $^{19}$F contained in PTFE. The detector responses to $^{137}$Cs (Columbia) and $^{133}$Ba (Case) show only the gamma ray band.

\begin{figure}[htbp]
\includegraphics[height=5.4cm]{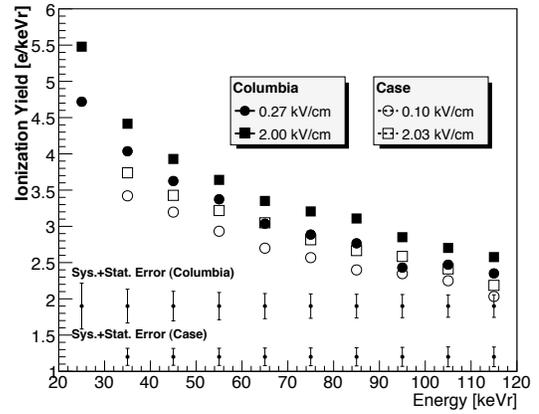}
\caption{\label{NRIY:Combine} Energy dependence of nuclear recoil ionization yield at different drift fields.}
\end{figure}

\textbf{Nuclear Recoil Ionization Yield:} 
The ionization yield is defined as the number of observed electrons
per unit recoil energy ($\rm e^-/keVr$).  The nuclear recoil
ionization yield as a function of energy is shown in
Fig. \ref{NRIY:Combine}, for several drift fields. The uncertainty on
the yield is dominated by the systematic error from the $^{57}$Co S2
calibration. 

Fig. \ref{yield_vs_field} summarizes the  relative light and charge yields as a function of drift field for different particles in LXe: 122~keV gamma rays from \cite{Aprile:2006chargelight,Shutt:2006calibration}, 56~keV Xe nuclear recoils from \cite{Aprile:2005xenonscint} and this paper, and 5.5~\&~5.3~MeV alphas from \cite{Aprile:1991alphas} and the Case detector.  The relative light yield $S(E)/S_0$ is the light yield relative to that at zero field $S_0$. The relative charge yield $Q(E)/Q_0$ is the charge collected relative to that at infinite field (\emph{i.e.}, with no recombination) $Q_0$. For nuclear recoils, $Q_0=E_r \cdot \mathcal{L}/W_e$, where $E_r$ is the recoil energy, $\mathcal{L}$ is the supression predicted by Lindhard theory \cite{Lindhard:1963nuclearsupression}, and $W_e = 15.6$ eV \cite{Takahashi:1975Winxenon} is the average energy required to produce an electron-ion pair in LXe. For gammas and alpha particles, we divide the known energy by $W_e$ to obtain $Q_0$. 

\textbf{Discussion:}
The important characteristics of the measured ionization yield of nuclear recoils in LXe are its value relative to that of other particles (Fig. \ref{yield_vs_field}), its energy dependence (Fig. \ref{NRIY:Combine}), and its field dependence (Fig. \ref{yield_vs_field}).  The yield is a function of the Lindhard factor and the amount of recombination in the track.   The Lindhard factor has no field dependence,  and does not  affect the relative charge yield (Fig. \ref{yield_vs_field}).  Lindhard does predict a slight decrease in charge yield with decreasing energy, the opposite of what we see.  Therefore we expect recombination and its dependence on track density and shape to explain the above phenomena.

\begin{figure}
\includegraphics[height=4.9cm]{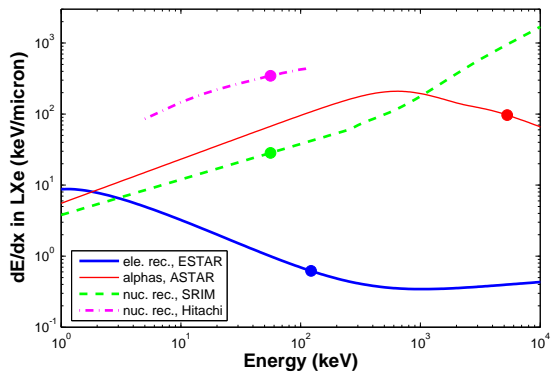}
\caption{\label{xetest:dedx}Predicted electronic stopping power, dE/dx, for different particles in LXe.  The circles indicate the particle energies discussed in the text.}
\end{figure}

Recombination is primarily a function of electric field and ionization
density, with stronger recombination at low fields and in denser
tracks.  Ionization density along a track corresponds roughly to
electronic stopping power, plotted in Fig. \ref{xetest:dedx} for
alphas, electrons, and Xe nuclei in LXe, as given by ASTAR, ESTAR, and
SRIM \cite{Website}, respectively .  Also shown is the total energy lost to electronic excitation per path-length for Xe nuclei, which differs from the electronic stopping power in that it includes energy lost via electronic stopping of secondary recoils \cite{Hitachi:2006}. 

At lower energies, the stopping power for Xe nuclei decreases, indicating lower recombination and higher ionization yield,  as observed.  The stopping power, including daughter recoils for Xe nuclei at 56~keV is higher than that of alphas at 5.5~MeV.  This is in conflict with what we observe in the relative ionization yields, indicating that the higher ionization yield of a low energy Xe recoil compared to an alpha is due to the different track geometries  for these two particles. Recombination for alphas  is enhanced because of their cylindrical track shape with a very dense core and a "penumbra" of delta rays.  SRIM simulations show that nuclear recoil tracks have many branches, with most energy lost by  secondary recoils. Each of these secondary branches presumably end in a very sparse track since the stopping power falls with energy.  This may result in a low-density halo of ionization around the track, which does not recombine even at very low fields.  
 If this lack of recombination extends to zero field, it would explain
 the  difference between the measured scintillation yield for nuclear
 recoils and  the Lindhard prediction. The importance of the track
 geometries is also seen in the striking contrast between the field
 dependence of nuclear recoils and those of electrons and alphas
 (bottom graph of Fig. \ref{yield_vs_field}), though electrons have a
 much lower, and alphas a comparable stopping power to nuclear
 recoils. Our new finding of the weak field dependence for nuclear
 recoils in LXe has the important practical consequence for dark
 matter detection that  only modest fields are needed for the
 background-discriminating measurement of nuclear recoils. 

\textbf{Background Discrimination:}
The different values of S2/S1 between nuclear recoils and electron
recoils make liquid xenon an excellent target to discriminate between
these two types of interaction.  Acceptance windows for nuclear
recoils were defined from the neutron elastic recoil band as shown in
Fig. \ref{2Dplots} and \ref{2Dcase}.  The rejection of background
electron recoils was measured by irradiating both detectors with pure
gamma ray sources ($^{133}$Ba or $^{137}$Cs), which produce low energy
electron recoils from Compton scatters, and recording the number of
electron recoils outside of 50\% acceptance nuclear recoil windows.
The Columbia detector, with higher S1 light collection, achieved an
electron recoil rejection efficiency of 98.5\% independent of energy
down to 20 keVr at 2 kV/cm drift field.  This efficiency appears to be
limited by charge loss at the edge of the detector where the field was
non-uniform (no field shaping rings were used in this prototype).  The
rejection efficiency in the Case detector improved with increasing
energy (because of the  non-optimal light collection) and was better
than 99.5\% above 80keVr, presumably due to a better field geometry
with closer electrodes  and better S2 resolution.  The rejection
efficiency improves with higher drift field (in the Case detector,
e.g., from 91\% at 0.2 kV/cm to 97\% at 4.6 kV/cm for 35 keVr). With
an optimized field configuration, and with the XY position sensitivity
of a LXeTPC as proposed for the XENON experiment
\cite{Aprile:2002xenon}, the problem of edge events should be
negligible. The dual phase xenon technique thus appears very promising
for a large-scale detector with powerful electron recoil background
discrimination for a sensitive WIMP dark matter search. 
	
The authors would like to thank A. Hitachi for insight and discussion on particle interaction in LXe.  TS would like to thank K. McDonald and C. Lu for advice and support.  This work was funded by NSF grants PHY-03-02646 and PHY-04-00596.


\end{document}